\documentclass[Afour,sageh,times]{sagej}

\usepackage{moreverb,url}

\usepackage[colorlinks,bookmarksopen,bookmarksnumbered,citecolor=red,urlcolor=red]{hyperref}

\newcommand\BibTeX{{\rmfamily B\kern-.05em \textsc{i\kern-.025em b}\kern-.08em
T\kern-.1667em\lower.7ex\hbox{E}\kern-.125emX}}

\usepackage{listings}
\usepackage{lipsum}
\usepackage{graphicx}
\usepackage{flushend}

\newcommand\revised{}

\begin{document}

\runninghead{Gray and Stratford}

\title{A Lightweight Approach to Performance Portability with targetDP}

\author{Alan Gray\affilnum{1} and Kevin Stratford\affilnum{1}}

\affiliation{\affilnum{1}EPCC, The University of Edinburgh}

\corrauth{Alan Gray, 
EPCC, The University of Edinburgh\\
Edinburgh, EH9 3FD, UK}

\email{a.gray@ed.ac.uk}

\begin{abstract}
Leading HPC systems achieve their status through use of highly
parallel devices such as NVIDIA GPUs or Intel Xeon Phi many-core
CPUs. The concept of performance portability across such
architectures, as well as traditional CPUs, is vital for the
application programmer. In this paper we describe targetDP, a
lightweight abstraction layer which allows grid-based applications to
target data parallel hardware in a platform agnostic manner. We
demonstrate the effectiveness of our pragmatic approach by presenting
performance results for a complex fluid application (with which the
model was co-designed), plus a separate lattice QCD particle physics
code. For each application, a single source code base is seen to
achieve portable performance, as assessed within the context of the
Roofline model. TargetDP can be combined with MPI to allow use on
systems containing multiple nodes: we demonstrate this through
provision of scaling results on traditional and GPU-accelerated large
scale supercomputers.
\end{abstract}


\maketitle

\setcounter{secnumdepth}{4}

\section{Introduction}

Each new generation of HPC system must offer an increase in the number
of operations that can be performed per second. Performance advances
in hardware are realized through parallelism, where the trend is not
just to increase the number of compute nodes but also to increase the
capability of each node through use of highly parallel devices such as
Graphics Processing Units (GPUs) or many-core CPUs. The key challenge
for application programmers is how to achieve ``performance
portability'', such that a single, easily maintainable source code
base can run with optimal performance across the range of modern
parallel architectures.

Most HPC applications have, over decades of development targeting
traditional CPU-based systems, grown to many thousands of lines of
code written in base languages such as C, C\texttt{++} and Fortran.
Compiler technology cannot automatically transform such raw code into
executables that can always perform well across such architectures, so
some form of modernization is required at the source code level. In
order to achieve performance portability, it is inevitable that the
concept of abstraction must play a role in one form or another, to
allow generic syntax to be mapped to a range of different
executables. There are two potential pitfalls with this approach,
though. The first is that any code-generation software required to
perform the mapping must be maintained along with the application, so
it is important that the software sustainability problem is not just
shifted from the application into the generator. It is very difficult
to avoid this problem for any sophisticated code-generation
mechanism. If a large community can reach agreement, then the
maintenance can be shared, but of course even when achievable this can
take time. The second potential pitfall is that the modernization
necessary can be relatively disruptive, which can discourage
developers from committing to the new approach.

In this paper we address these problems in a pragmatic way, where our
abstraction is intentionally kept as simple and lightweight as
possible. We demonstrate the effectiveness of this approach through
performance portability results for real complex applications. The
targetDP model \citep{gray2014targetdp} provides an abstraction layer
which allows applications to target Data Parallel hardware in a
platform agnostic manner, by abstracting the memory spaces and
hierarchy of hardware parallelism. Applications written using targetDP
syntax are performance portable: the same source code can be compiled
for different targets (where we currently support NVIDIA GPU
accelerators and modern multicore and manycore CPUs including the
Intel Xeon Phi which rely on vectorization). TargetDP is implemented
using a minimal set of C-preprocessor macros and library functions, so
is easily maintainable, sustainable and extendable. There are no
special datatypes, so it is possible to adapt incrementally and also
inter-operate with other programming paradigms. The model is
appropriate for abstracting the parallelism contained within each
compute node, and can be combined with, e.g. MPI to allow use on
systems containing multiple nodes.

The model and implementations were developed in co-design with a
particular HPC application, Ludwig \citep{desplat2001ludwig}, which
uses lattice Boltzmann and finite difference techniques to simulate a
wide range of complex fluids. However, we demonstrate general
applicability using the separate MILC code \citep{MILC}, which
simulates fundamental particle physics phenomena through Lattice QCD
techniques. TargetDP is particularly suited to the structured grid
techniques employed by both of these applications since the regularity
and parallelism can be exploited in a straightforward
manner. Therefore, whilst at present the model can be thought of as
being specific to this computational domain, it should also applicable
to a wider class of data parallel problems.

In Section \ref{sec:background} we describe these applications,
focusing on the test cases used for performance results, and we also
discuss other approaches to performance portability. In Section
\ref{sec:targetDP}, we present the abstractions used to achieve
performance portability. We go on to present single-processor results
in Section \ref{sec:performance}, where we analyse performance across
architectures following the Roofline model \citep{williams2009roofline}
(which gives standard guidelines for comparing against the capability
of the hardware). We extend the analysis to include 
scaling over multiple nodes of supercomputers in Section
\ref{sec:scaling}, through the combination of targetDP and MPI.

\section{Background}\label{sec:background}

\subsection{The Applications}
Many scientific and engineering problems are made tractable through
the discretization of space and time, to allow representation and
evolution within a computer simulation.  In this paper we focus on two
such applications, Ludwig and MILC (both of which are open source, use
C as a base language and MPI for node-level parallelism).

\subsubsection{Ludwig}

\mbox{ }\\ Many substances which are neither solids nor simple fluids can be
classed as ``soft matter'', also known as ``complex fluids''. Familiar
examples include foodstuffs, lubricants, cosmetic and health-care
items, bodily fluids (such as those found in the joints) and,
pertinent to this paper, liquid crystals (LCs). The versatile Ludwig
simulation package \citep{desplat2001ludwig} is a vital link between
theory and experiment in enabling new and improved soft matter
materials. Like many such applications, Ludwig represents space as a
3D structured grid (or {\it lattice}), where the physical system, at a
certain point in time, is represented by a set of double precision
values at each lattice point. The lattice Boltzmann (LB) method
\citep{succi2001lattice} is used to evolve the hydrodynamics in a
standard timestepping manner.  Ludwig, however, can simulate not just
simple but complex fluids; the hydrodynamical evolution is coupled
with other finite difference techniques to properly represent the
substance under investigation.

LCs are well known for their utility in displays, but are also
prevalent in a variety of other technological items and natural
systems. There is still much to be understood about the range of
possible LC systems, which are a key focus of interest in current
research using coarse-grained
methods \citep{henrich2013rheology}\citep{tiribocchi2014switching}. The
LC simulation, which comprises one of the testcases used later in this
paper, couples an ``order parameter'' field (a $3\times 3$ tensor,
which is symmetric and traceless) representing the composition and
structure of the crystal, and a ``distribution'' field representing
the flow of the fluid. The former is evolved via an
advection-diffusion equation appropriate for rod-like molecules, and
the latter via LB. They interact through a local force, derived from
the former, which acts on the latter at every timestep of the
simulation. The main computational components of each timestep are as
follows, where the names given in quotes are those used later in this
paper when presenting performance results. The ``Collision'' models the
interaction of the individual fluid molecules and the ``Propagation'', which mimics the convection of the fluid, involves displacing
the fluid data one lattice spacing in the appropriate direction. The
force which couples the order parameter with the distribution is
calculated as the divergence of the ``Chemical stress'', which in turn
is calculated as a function of the order parameter field and its
``Order Parameter Gradients'' derivatives. The ``LC Update'' involves
evolution of the order parameter itself, using a finite difference
implementation of the Beris-Edwards model with the Landau-de Gennes
free energy functional
\citep{beris1994thermodynamics}\citep{prost1995physics}. The
``Advection'' involves calculating the flux in the order parameter due
to the advective bulk flow, and ``Advection Boundaries'' includes the
effects of any boundary conditions to this.

To allow utilization of multi-node computing architectures, Ludwig is
parallelized using domain decomposition and MPI in a standard way.
``Propagation'', ``Advection'', ``Advection Boundaries'', and ``Order Parameter
Gradients'' involve updates based on neighbouring lattice site data, so
are classed as stencil operations, whereas ``Collision'', ``Chemical Stress''
and ``LC Update'' involve operations local to each lattice site.  To allow
for the former, each local sub-domain is surrounded by a halo region
populated using neighboring sub-domain data through MPI
communications. In recent years we have adapted Ludwig for use on
multiple GPUs in parallel (as described in
\citep{bfd6b9c81b4d47ce8fa5e97992f49c44}, \citep{gray2015scaling} and
\citep{534423c8fcbf4ac6b7e0b95118161e9c}), which has enabled an
improved understanding of the interactions between LCs and large
colloidal particles \citep{stratford2015large}.

\subsubsection{MILC}
\mbox{ }\\ Matter consists of atoms, which in turn consist of nuclei and
electrons. The nuclei consist of neutrons and protons, which comprise
quarks bound together by gluons. The theory of how quarks and gluons
interact to form nucleons and other elementary particles is called
Quantum Chromo Dynamics (QCD). For most problems of interest, it is
not possible to solve QCD analytically, and instead numerical
simulations must be performed. Such “Lattice QCD” calculations are
known to give excellent agreement with experiment
\citep{davies2004high}, at the expense of being very computationally
intensive.

The MIMD Lattice Computation (MILC) code is a freely available suite
for performing Lattice QCD simulations, developed over many years by a
collaboration of researchers based in the US \citep{MILC}. The test
case used in this paper is derived from the MILC code (v6), and forms
a component Unified European Application Benchmark Suite (UEABS), a
set of application codes designed to be representative of HPC usage in
the European Union \citep{UEABS}. This consists of the inversion of the
Wilson Dirac operator using the conjugate gradient method. The work
presented in this paper is part of a project to extend the UEABS suite
to accelerators.

Spacetime is discretized into a 4D grid, and at each point on the grid
exists a set of values that represent the quark and gluon content at
that point. These structures are relatively small (compared to the
size of the grid) complex vectors and matrices, and the conjugate
gradient algorithm involves correspondingly small linear algebraic
operations at each point on the grid. With reference to the labels
used for performance analysis later in this paper, ``Extract''
involves extracting the quark field from one representation to
another, and ``Extract and Mult.'' involves a similar quark field
manipulation together with a matrix-vector multiplication to interact
with the gluon field. The ``Insert and Mult'' and ``Insert'' parts
involve the reverse process. ``Scalar Mult. Add'' involves a scalar
multiplication and vector addition for the quark field. These
operations are all local to each grid point, but the ``Shift''
operation involves shifting the quark field data in each direction,
which is a stencil operation. When running on a single processor, this
just involves local memory copy operations but, when operating with
node-level parallelism, MPI is used within this operation to buffer
and move the data that must cross subdomain boundaries.

\subsection{Other Approaches to Performance Portability}\label{sec:perfport}

The use of a low-level native programming model, e.g. CUDA for NVIDIA
GPUs, permits excellent performance on the native architecture but lacks
portability. 

The OpenCL standard by the Khronos group \citep{opencl} gives a similar
model to CUDA but with improved portability, particularly to AMD GPUs,
and in principle to X86 CPUs including Xeon Phi.  Intel do not
currently plan to support OpenCL, however, on the next generation of
Xeon Phi (Knights Landing). It is also relatively low level and
complex, and can have performance overheads relative to native
approaches (although the situation is improving
\citep{mcintosh2014performance}). The SYCL standard \citep{sycl}, also
by Khronos, has recently emerged: this aims to be higher level and
more productive, particularly for C++ applications. There also is
active research into automatic generation of hardware-specific OpenCL
code, e.g. with functional programming techniques and rewrite rules
\citep{steuwer2015generating}.  The use of OpenCL, within the types of
applications discussed in this paper, would mean increased complexity
and compromised portability and performance, compared with
targetDP. 

\revised{ The OpenACC directive-based standard \citep{openACC} is
  designed to provide a higher-level and more user-friendly
  alternative to CUDA/OpenCL for development of GPU-accelerated
  applications. The idea is that a greater onus is put on the compiler
  to automatically manage data movement and computational offloading,
  with help from user-provided directives. The model is prescriptive,
  such that the compiler is free to map parallelism and data movement
  to the specific hardware in an automatic manner (although a range of
  directives and clauses are available providing the user with the
  ability to override.) OpenACC offers portability across NVIDIA and
  AMD GPUs, and has recently expanded support to multi-core
  CPUs. However, there is still no support for Intel Xeon Phi, and
  Intel do not participate on the OpenACC committee. Whilst OpenACC is
  not yet a viable option for true performance portability, it will be
  interesting to monitor progress. A key concern is whether the level
  of flexibility available is adequate for the intricacies associated
  with real complex applications (e.g. the interaction of particles
  with liquid crystals in Ludwig which requires very fine-grained
  control over movement of data between memory spaces to ensure good
  performance). Similarly to OpenACC, The OpenMP directives-based
  standard \citep{openmp} has, since version 4.0, included support for
  accelerators. OpenMP aims to be more portable than OpenACC,
  including participation from Intel and support for the Xeon
  Phi. However, the required directives actually differ across
  architectures, so the approach ultimately lacks
  portability. Similarly to OpenACC, the relatively high-level OpenMP
  model offers productivity at the expense of user control (and hence
  often performance).  }

Alternatively, a separate source-to-source generator (e.g. a
preprocessor, compiler or library) can be used to transform generic
appropriately defined syntax into multiple forms, which can each be
compiled by the standard compiler for the specific target
architecture. Kokkos \citep{edwards2012manycore} and RAJA
\citep{hornung2014raja}, both C++ abstraction frameworks, have
conceptual similarities to the work presented in this paper, since they
provide suitable abstractions for node-level parallelism and data
movement to allow portable performance across modern architectures.
These are much more sophisticated than targetDP since they aim to be general
purpose models and support a wide range of application areas. The
caveat, however, is the disruption required to adapt legacy
applications (particularly those which do not already use
C++). Additionally, users are often hesitant to commit to third-party
models and libraries which do not form any standard so have no
guaranteed longevity. The developers are addressing the latter
by lobbying for much of the models to be integrated into the C++
standard. The work described in this paper, conversely, exploits the
domain specificity (to grid-based codes) to retain simplicity, so
these issues are much less severe. Other frameworks such as OCCA
\citep{medina2014occa} and HEMI \citep{HEMI} are alternative C++ based
models, which have varying levels of support for explicitly targeting
SIMD units.

Legion \citep{Bauer:2012:LEL:2388996.2389086} is another very
sophisticated model. This not only provides portable abstractions, but
also allows the programmer to express tasks and dependencies which,
through a runtime system, can be automatically executed with optimal
ordering, dependent on the hardware resource in use. This is very
powerful in extracting performance, but even more disruptive to the
programmer that the approaches discussed above. Legion is really
designed as a low level interface for other domain specific languages
to be built on. Recently, however, the Regent compiler
\citep{slaughter2015regent} has been introduced which generates Legion
code from a higher-level model.

So clearly, the best option for a programmer wishing to invest in code
modernization depends on the application.  Given our pragmatic and
simple approach, one aim of this paper is to clearly describe, in
an accessible manner, those concepts which are common, such as the
abstraction of memory accesses and data parallel operations, the
expression of data locality, and the resulting performance portability
achieved. We hope this is of interest even to those readers who decide
to follow one of the other available development routes.

\section{targetDP}\label{sec:targetDP}

In this section we describe the targetDP programming model (first
introduced in \citep{gray2014targetdp}). The aim is to provide, in the
simplest possible way, an abstraction that can allow the same
data-parallel source code to achieve optimal performance on both
CPU-like architectures (including Xeon Phi) and GPUs.  Before
describing our model, we first give details on how we abstract memory
accesses through a simple layer. This is necessary to allow
architecture-specific data-layouts, which are vital for memory
bandwidth performance.


\subsection{Data Layout}

To facilitate optimal memory bandwidth, it is crucial that the layout
of data in memory is tailored for the specific memory access patterns
that result from the execution of a targetDP application on any
specific architecture.

For many grid-based applications, the simulation data is
``multi-valued'': comprised from multiple numerical values located at
each point on the grid, and how we choose to store this in memory can
have a dramatic effect on performance. To illustrate the available
options, a useful analogy is that of red (r), green (g) and blue (b)
values collectively representing each pixel of an image. For brevity,
let's assume that our image is very small at only 4 pixels. The naive
way to store the entire image is \textbar rgb\textbar rgb\textbar
rgb\textbar rgb\textbar. This is known as the Array of Structures
(AoS) format since we are storing the rgb values consecutively as each
entry of an array of pixels. An alternative is the
Structure of Arrays (SoA) storage scheme \textbar rrrr\textbar
gggg\textbar bbbb\textbar. These are the two opposite extremes of a more
general scheme: the Array of Structures of (short) Arrays (AoSoA)
format, where the length of the short array can vary. For example,
\textbar \textbar rr\textbar gg\textbar bb\textbar \textbar \textbar
rr\textbar gg\textbar bb\textbar \textbar , which has a short array
length of 2, is a layout in which we store the first 2 values of each
of red, green and blue, then repeat for the next 2 of each. The AoS
and SoA formats described above have a short array length of 1 and 4
respectively in this case.

To represent such multi-dimensional structures in C or C++,
it is usually most straightforward to use 1-dimensional arrays in the application
and to ``flatten'' accesses through linearization of the indices. Let
\verb+ipix+ and \verb+irgb+ denote the pixel and rgb value indices of
this 2-D problem respectively, and \verb+N+ be the total number of
pixels (with, of course, 3 being the total number of rgb components). Then, if using the AoS scheme we access the
image with code such as \verb| image[ipix*N+irgb]|. The alternative
SoA scheme would instead use \verb|image[irgb*3+ipix]|, and AoSoA
would use 

\scriptsize
\begin{verbatim}
image[((ipix/SAL)*3*SAL+irgb*SAL+(ipix-(ipix/SAL)*SAL))] 
\end{verbatim}

\normalsize
\noindent (where the short array length is specified through the \verb+SAL+ preprocessor-defined integer constant).

For most lattice-based scientific simulations, including those
discussed in this paper, the main bottleneck is memory bandwidth.  The
specific layout of data in memory has a major effect on memory
bandwidth performance and unfortunately the best layout differs across
architectures. Therefore, for performance portability it is required
memory accesses should not be explicitly written, as above, within the
application. Instead they should be abstracted through an intermediate
layer, in which the optimal layout can be specified through a
configuration option without any changes to the application itself.

Following the philosophy of this paper, we achieve this in a 
simple way.  We define macros using the C-preprocessor
to provide the required memory addressing layer. At the application
level, our accesses become, e.g., \verb+image[INDEX(irgb,ipix)]+,
where different versions of the \verb+INDEX+ macro can expand to match
different linearizations (or, equivalently, the single generic AoSoA
version can be used with different short array lengths).

We can follow the practice described in this section for the
multi-dimensional data structures that naturally exist in real
structured grid problems, which have one dimension corresponding to
the grid points (c.f. the pixel dimension) and one or more dimensions
correspond to the components stored at each grid point (c.f. the rgb
values). Note that in practice we also pass extent values into the
macros to allow flexibility regarding grid dimensions.

\subsection{Data Parallel Abstractions}

The execution time of grid-based applications is typically dominated
by the parts of the code that perform data-parallel operations across
the grid: at each point on the grid, the multi-valued data resident at
that point is updated (sometimes using data from neighbouring grid
points). We use the the terminology ``host'' to refer to the CPU on
which the application is running, and ``target'' refers to the device
targeted for execution of these expensive operations.  When the target
is a GPU, there is clearly a distinction in hardware between the host
and target. But the host and target can also refer to the same
hardware device, for instance when we are using CPU (or Xeon Phi)
architectures. Importantly, we always retain the distinction of host
and target in the application, to allow a portable abstraction.

The aim of targetDP is to provide a lightweight abstraction that can
map data-parallel codes to either CUDA or OpenMP, whilst allowing good
vectorization on CPU-like architectures. \revised{Note that, since the
  abstraction layer maps directly to the underlying model in each case, by design there is negligible overhead 
  compared to an equivalent native implementation.}

\subsubsection{Thread-level Parallelism}
\mbox{ }\\ Consider a serial C code
which involves a data-parallel loop over \verb+N+ grid points, e.g.

\begin{minipage}{\linewidth}
\begin{lstlisting}[linewidth=\columnwidth,breaklines=true,basicstyle=\tt\footnotesize]
 for (index = 0; index < N; index++) { 
   ...
 }
\end{lstlisting}
\end{minipage}

\noindent The \verb+...+ refers to whatever operation is performed at each grid point.
An OpenMP implementation of this code would instruct the compiler to decompose the loop across threads with use of a pragma, e.g.

\begin{minipage}{\linewidth}
\begin{lstlisting}[linewidth=\columnwidth,breaklines=true,basicstyle=\tt\footnotesize]
 #pragma omp parallel for
 for (index = 0; index < N; index++) { 
   ...
 }
\end{lstlisting}
\end{minipage}

\noindent The thread-level parallelism in CUDA does not involve the concept of loops. Instead, the grid-based operation becomes a kernel to be executed in parallel by a number of threads, e.g.

\begin{minipage}{\linewidth}
\begin{lstlisting}[linewidth=\columnwidth,breaklines=true,basicstyle=\tt\footnotesize]
__global__ void scale(double* field) { 

  int index=blockDim.x*blockIdx.x+threadIdx.x;
  if (index < N)
  { 
    ...
  }
  return;
}
\end{lstlisting}
\end{minipage}

\noindent The operation must be contained within a function, where the \verb+__global__+ keyword specifies it should be compiled for the GPU, and CUDA internal variables are used to retrieve a unique index for each thread.

So, therefore, we can provide an abstraction that can map to either OpenMP or CUDA  by introducing the targetDP syntax:

\begin{minipage}{\linewidth}
\begin{lstlisting}[linewidth=\columnwidth,breaklines=true,basicstyle=\tt\footnotesize]
__targetEntry__ void scale(double* field) { 

  int index;
  __targetTLP__(index, N) { 
    ...
  }
  return;
}
\end{lstlisting}
\end{minipage}

\noindent \verb+__targetEntry__+ and \verb+__targetTLP__+ are
C-preprocessor macros. The former is used to
specify that this function is to be executed on the target and it will
be called directly from host code (where the analogous
\verb+__target__+ syntax is for functions called from the target). For
implementation in CUDA it is mapped to \verb+__global__+, so the
function becomes a CUDA kernel, and for the \verb+C+ implementation
the macro holds no value so the code reduces to a regular function. We
expose the grid-level parallelism as thread-level parallelism (TLP)
through the \verb+__targetTLP__(index,N)+ syntax, which is mapped to a
CUDA thread look-up (CUDA implementation), or an OpenMP parallel loop
(C implementation). In the former case, the targetDP example therefore
maps exactly to the CUDA example. For the latter, the only difference
to the OpenMP example is that the code is now contained in a function.

\subsubsection{Instruction-level Parallelism}
\mbox{ }\\ We must also be able to achieve good vectorization on CPU-like
architectures. The compiler will try to create vector instructions
from innermost loops, but the problem with the above is that the
extent of the innermost loop is dependent on the application, so does
not necessarily map well to any specific hardware vector length (and
furthermore not all such loops over multi-valued data are parallel in
nature). To give a simple example, imagine that \verb+...+ is given by

\begin{minipage}{\linewidth}
\scriptsize
\begin{lstlisting}[linewidth=\columnwidth,breaklines=true,basicstyle=\tt\footnotesize]
   int iDim;
   for (iDim = 0; iDim < 3; iDim++) 
      field[INDEX(iDim,index)] = 
          a*field[INDEX(iDim,index)];
\end{lstlisting}
\end{minipage}

\noindent The \verb+field+ data structure has 3 components at each
grid point, and an extent of 3 for the innermost loop is not optimal
for creating vector instructions of size, e.g. 4 or 8 for CPU or Xeon
Phi (which, when using double precision is the natural vector length
for these architectures which feature 256-bit AVX and 512-bit IMCI
instructions), respectively. The solution is that we can allow
striding within our implementation of \verb+__targetTLP__+, such that
each thread works, not on a single lattice site, but a chunk of
lattice sites. Then, at the innermost level, we can re-introduce this
grid-level parallelism for the compiler to vectorize.  The size of the
chunk (i.e. length of the stride), which we call the ``Virtual Vector
Length'' (VVL), can be tuned to the hardware through a targetDP
configuration option at compile time. The new instruction-level
parallelism (ILP) is specified using the \verb+__targetILP__+
syntax. When \verb+VVL>1+, then \verb+__targetILP__+ is mapped to a
short loop over the sites in the chunk, augmented with the OpenMP SIMD
directive. Since the loop extent, \verb+VVL+, appears as a constant to
the compiler (and is chosen to be a suitable value), vectorization is
straightforward. The ILP syntax, therefore, has an associated index
that ranges between \verb+0+ and \verb+VVL-1+, such that it can be
used as an offset when combined with the TLP thread index in accessing
arrays. The final targetDP implementation for this example is given in
Figure \ref{fig:targetDPkernel}. Note that we also include the
definition of the constant that is used in the multiplication: the
\verb+__targetConst__+ macro maps to \verb+__constant__+ in the CUDA
implementation (and holds no value for the C implementation) to allow
use of the constant memory cache on the GPU.

\begin{figure}[t]
\begin{lstlisting}[linewidth=\columnwidth,breaklines=true,basicstyle=\tt\footnotesize]
__targetConst__ double a; (*@\label{example:scalekernel:targetConst}@*) 

__targetEntry__ void scale(double* field) { (*@\label{example:scalekernel:targetEntry}@*) 

  int baseIndex;
  __targetTLP__(baseIndex, N) { (*@\label{example:scalekernel:targetTLP}@*) 

  int iDim, vecIndex;
  for (iDim = 0; iDim < 3; iDim++) {

    __targetILP__(vecIndex)  (*@\label{example:scalekernel:targetILP}@*) 
      field[INDEX(iDim,baseIndex+vecIndex)] =  
         a*field[INDEX(iDim,baseIndex+vecIndex)];       	  

  }
 }
 return;
}
\end{lstlisting}\vspace{-0.5cm}
\caption{A simple example of a targetDP kernel.}\label{fig:targetDPkernel}
\end{figure}

\subsubsection{Memory Management}
\mbox{ }\\ 

\begin{figure}
\begin{lstlisting}[linewidth=\columnwidth,breaklines=true,basicstyle=\tt\footnotesize]
 targetMalloc((void **) &t_field, datasize); (*@\label{example:scalelaunch:targetMalloc}@*) 
  
 copyToTarget(t_field, field, datasize); (*@\label{example:scalelaunch:copyToTarget}@*) 
 copyConstToTarget(&t_a, &a, sizeof(double)); (*@\label{example:scalelaunch:copyConstToTarget}@*) 
  
 scale __targetLaunch__(N) (t_field); (*@\label{example:scalelaunch:targetLaunch}@*) 
 targetSynchronize();(*@\label{example:scalelaunch:targetSynchronize}@*) 
  
 copyFromTarget(field, t_field, datasize); (*@\label{example:scalelaunch:copyFromTarget}@*) 

 targetFree(t_field); (*@\label{example:scalelaunch:targetFree}@*) 
\end{lstlisting}\vspace{-0.5cm}
\caption{The host code used to call the targetDP kernel.}\label{fig:targetDPcall}\vspace{-0.5cm}
\end{figure}

The \verb+scale+ function is called from host code as in Figure \ref{fig:targetDPcall}. This illustrates how the model draws a
distinction between the memory space accessed by the host and that
accessed by the target. The \verb+targetMalloc+ and \verb+targetFree+
API functions wrap \verb+cudaMalloc+ and \verb+cudaFree+ for the CUDA
implementation, and regular \verb+malloc+ and \verb+free+ for the C
implementation \footnote{\revised{Note that, as described in the Specification \citep{targetDPweb}, we also support targetCalloc which is analogous but for allocation with initialization to zero, and targetMallocUnified/targetCallocUnified which can take advantage of modern unified memory (although we do not use the latter in the benchmarks presented in this paper).}}. Similarly, the \verb+copy*+ routines abstract data
transfer in a portable fashion. It is assumed that code executed on the
host always accesses the host memory space, and code executed on the
target (i.e. within target functions) always accesses the target
memory space. The host memory space can be initialized using regular
\verb|C/C++| functionality (not shown here). For each data-parallel
data structure, the programmer should create both host and target
versions, and should update these from each other as and when
required.  Data parallel operations are achieved through passing target
pointers into target functions. The \verb+__targetLaunch__+ macro maps
to the regular CUDA launch syntax in the CUDA implementation, and
holds no value in the C implementation, and similarly for
\verb+targetSyncronize+.

For the C implementation, the host and target versions of data
structures will exist in the same physical memory space. The host
implementation can implement these as physically distinct, or can
instead use pointer aliasing: the model is agnostic to this. Note that
this explicit memory model will map well to future CPU-like
architectures, including the next generation of Xeon Phi, ``Knights
Landing'', which features a high-bandwidth stacked memory space in
addition to regular DRAM.

In applications, it is often necessary to perform reductions, where
multiple data values are combined in a certain way. For example,
values existing on each grid point may be summed into a single total
value. The targetDP model supports such operations in a simple but
effective way. It is the responsibility of the application to create
the array of values (using standard targetDP functionality) to act as
the input to the reduction operation. The application can then pass
this array to the API function corresponding to the desired reduction
operation (e.g. \verb+targetDoubleSum+ for the summation of an array
of double-precision values). If the required reduction operation does
not yet exist, the user can simply extend the targetDP API using
existing functionality as a template. 

In this section we have illustrated the basic key
functionality. Obviously, real applications are much more complex that
the example given above, and correspondingly there exists a range of
additional targetDP functionality, as detailed in the Specification
\citep{targetDPweb}. In particular, we provide functions that allow optimal
performance when data transfers between host and target can be
restricted to specific subsets of grid points.

\section{Performance}\label{sec:performance}

\begin{table}
\centering
\footnotesize
\resizebox{\columnwidth}{!}{%
\begin{tabular}{|c|c|c|c|} \hline
Processor & Product Details& Peak Perf.& Stream Triad \\   
 & & (Dbl. Prec.) & (Measured) \\  \hline  \hline 
Ivy-Bridge & Intel Xeon E5-2697 v2&259 Gflops& 49.8 GB/s \\ 
           & 12-core CPU @ 2.70GHz&&\\  \hline 
Haswell & Intel Xeon E5-2630 v3&154 Gflops&40.9 GB/s\\  
           & 8-core CPU @ 2.40GHz&&\\  \hline 
Interlagos&AMD Opteron 6274&141 Gflops& 32.4 GB/s\\  
          &16-core CPU @ 2.20GHz&&      \\  \hline 
Xeon Phi& Intel Xeon Phi 5110P  &1.01 Tflops&158.4 GB/s\\  
        & 60-core CPU @ 1.053GHz &  &     \\  \hline 
K20X & Nvidia Tesla K20X GPU &1.31 Tflops& 181.3 GB/s\\  \hline
K40&Nvidia Tesla K40 GPU&1.43 Tflops& 192.1 GB/s\\  
\hline\end{tabular}
}
\normalsize
\caption{Technical details of the processors used in the performance analysis.}\label{tab:procdetails} \vspace{-0.7cm}
\end{table}

\begin{figure*}[!t]
\centering
\includegraphics[width=7cm]{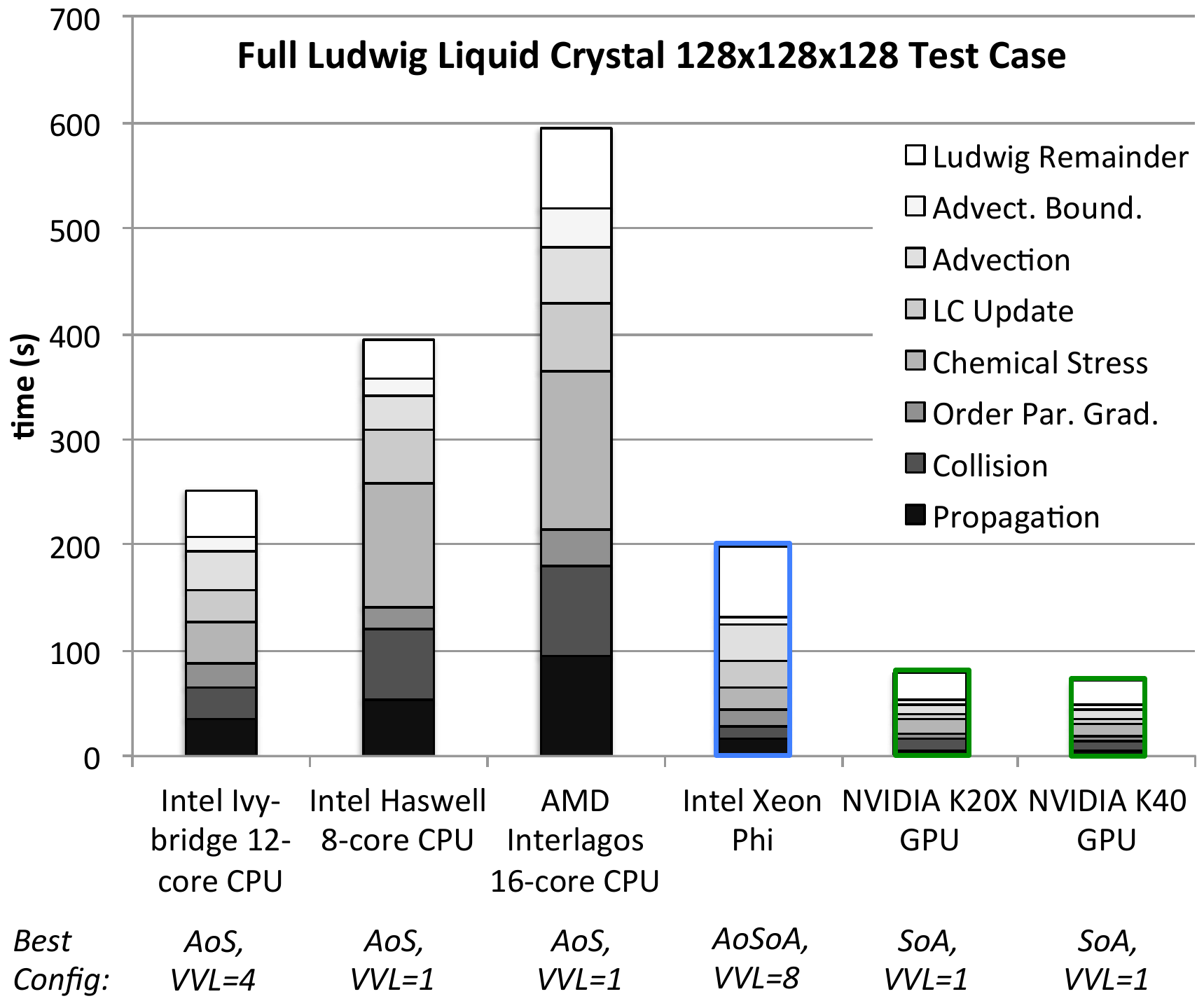}\hspace{1cm}
\includegraphics[width=7cm]{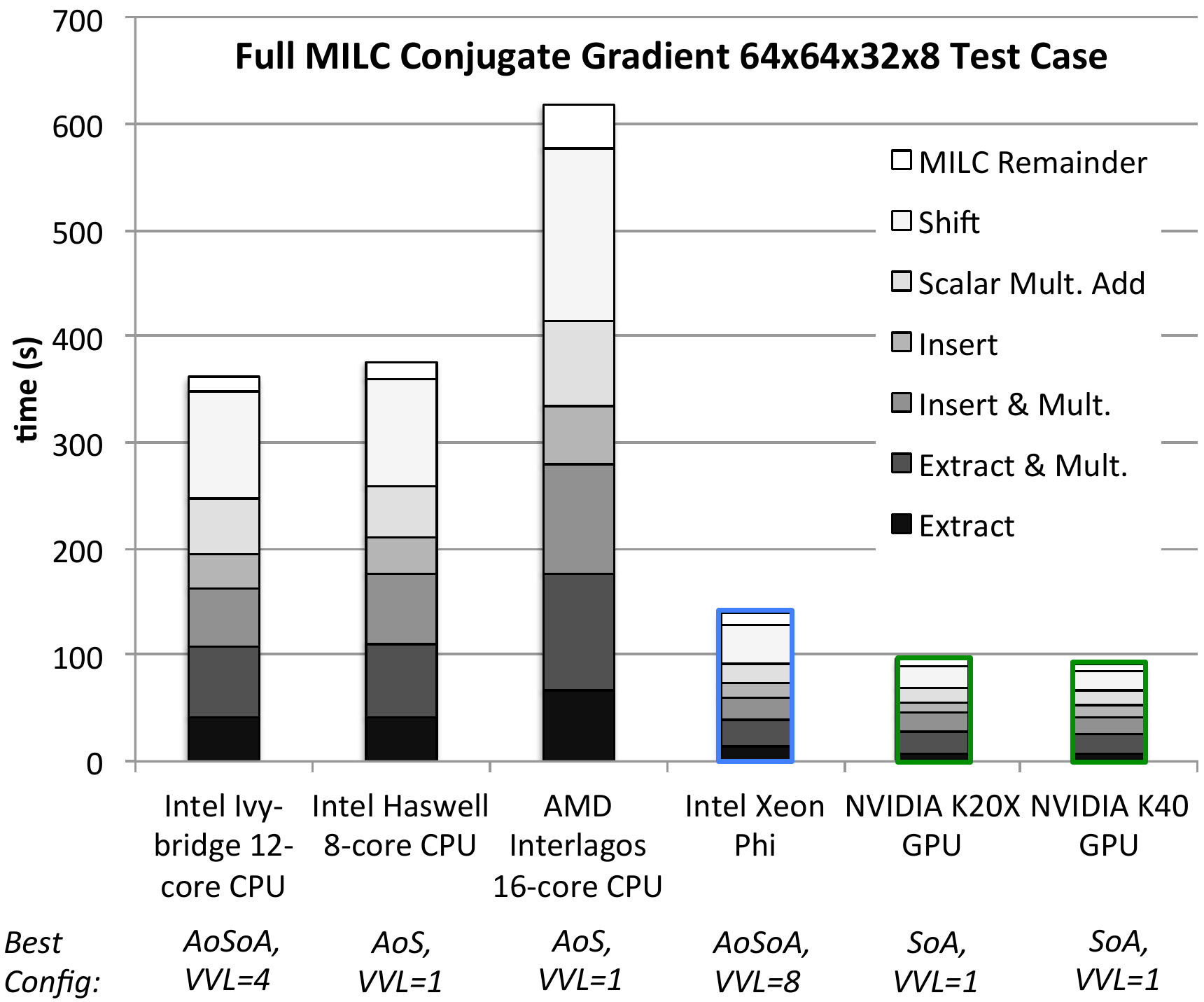}\\\vspace{0.5cm}
\includegraphics[width=11cm]{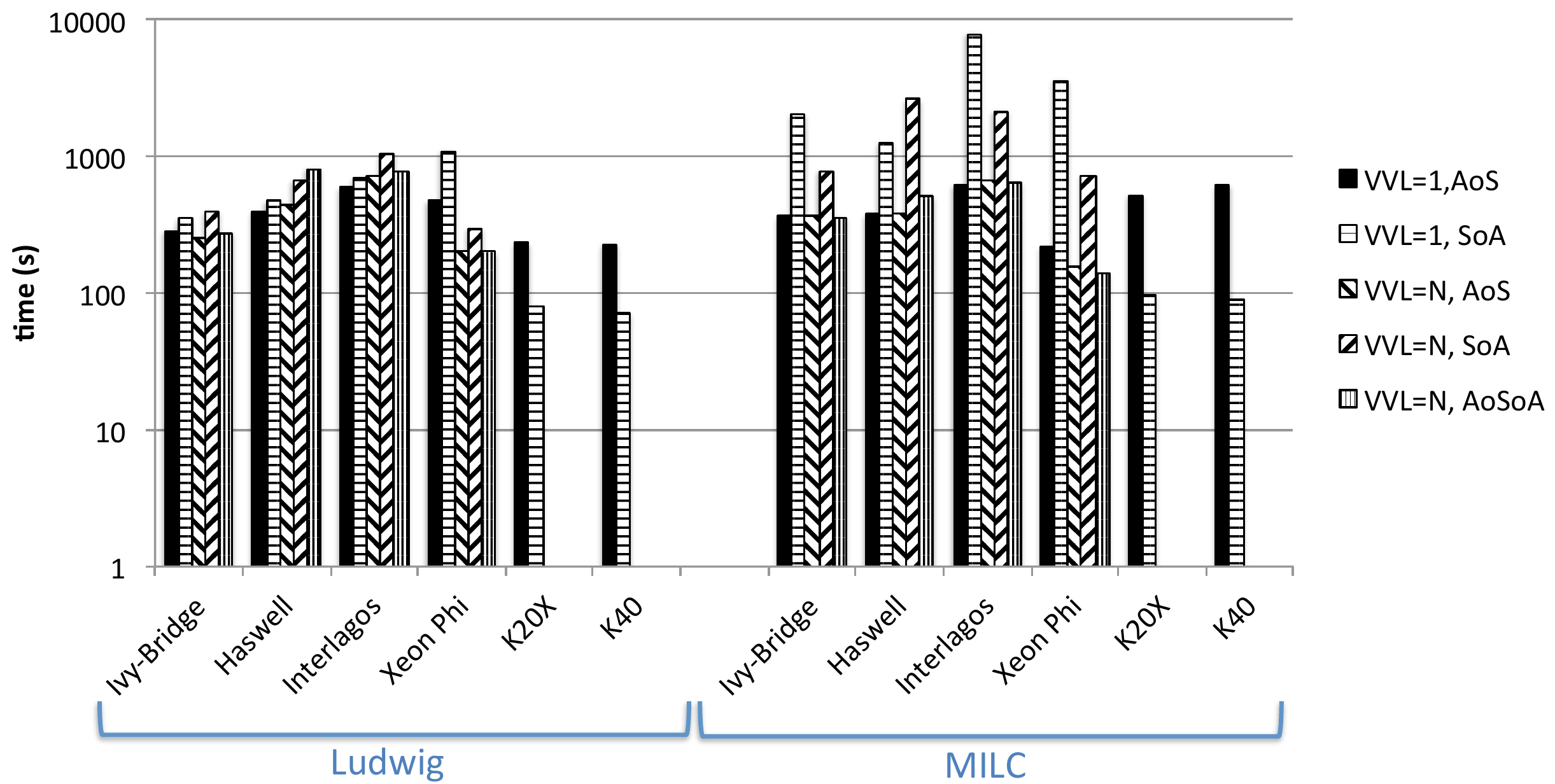} 
\vspace{-0.3cm}\caption{The time taken by the full Ludwig Liquid Crystal (top left) and MILC (top right) test cases on CPU, GPU and Xeon Phi architectures, where timings are decomposed into the different parts of the code. The same targetDP source code is used in each case and the best data layout and explicit vectorization configuration settings are shown, where variation across these options is shown in the bottom graph.}  
\label{fig:singleproc}
\end{figure*}

\begin{figure*}[t]
\centering
\includegraphics[width=11cm]{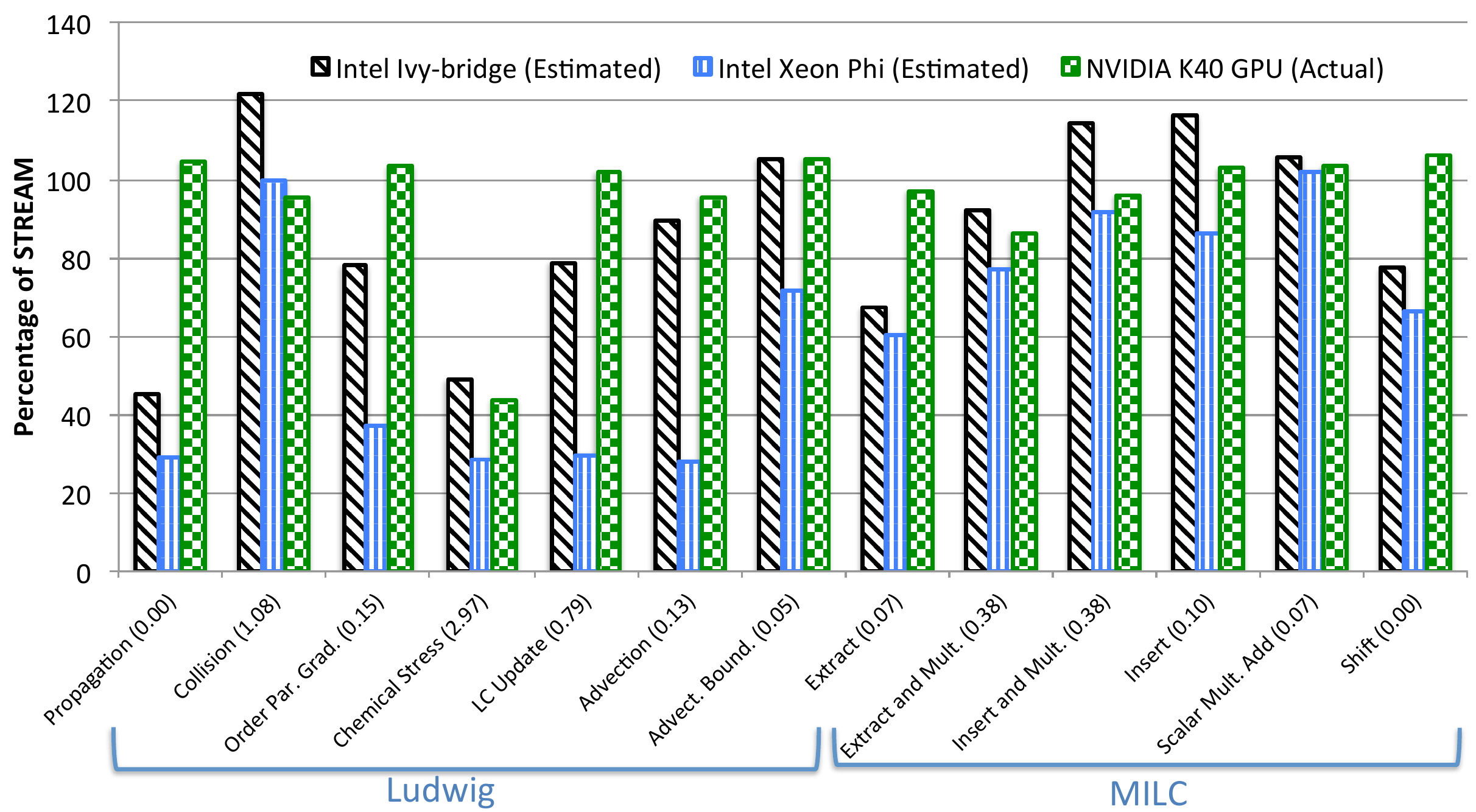} 
\vspace{-0.3cm}\caption{The memory bandwidth obtained by each Ludwig and MILC kernel, presented as a percentage of that obtained by the STREAM triad benchmark, across CPU, Xeon Phi and GPU architectures. For the application kernels, the GPU results are obtained using the NVIDIA profiler, and the other results are estimated from these using ratios of kernel time results to scale. The operational intensity of each kernel is given in brackets after the kernel name.}
\label{fig:bandwidth}
\end{figure*}


In this section, our main aim is to demonstrate that use of targetDP
allows, for each application, a single version of source code to be
performance portable across the range of relevant modern hardware
architectures. We first present the performance, for each of the
Ludwig and MILC codes, across different processors. We then go on
to compare these observations to hardware capability, using the
Roofline model to guide us.

Details of the processors used are given in Table
\ref{tab:procdetails}. We include regular multi-core CPUs, the Intel
Xeon Phi manycore CPU and NVIDIA GPUs. This choice naturally includes
those processors present in the supercomputers on which we will extend
our analysis in Section \ref{sec:scaling}: Ivy-bridge, Interlagos and
K20X. Haswell and K40 are newer lines than Ivy-Bridge and K20X
respectively, so are included give assurance that our results are
representative of modern technology.

To permit a clear comparison of the capabilities of the different
processor architectures we restrict our analysis, in this section, to
that using a single processor, be that a CPU or GPU. This means that,
for example, we will compare a single Ivy bridge CPU with a single
K40, where each is fully utilized. In section \ref{sec:scaling}, where
extend our analysis to multi-node systems, we will compare on a
node-by-node basis, with each node fully utilized, to give a
comparison which is more natural considering normal utilization of such
systems. 

For the CPU and Xeon Phi (where the latter is in the ``native'' mode
of use so can be treated on the same footing as a regular CPU) we use
the C implementation of targetDP, which maps to OpenMP (with the
number of threads set to best utilize the number of cores, or virtual
hyperthreaded cores if the latter is beneficial) in each case. Note
that the 16-core Interlagos processor is actually comprised from two
8-core chips in the same socket, so has two NUMA regions. Having
tested the different options of mapping to this architecture, we use
the best performing configuration of two MPI tasks each controlling 8
OpenMP threads (restricted to their respective NUMA regions). We use
the CUDA implementation of targetDP for GPU results, with the number
of CUDA threads per block set at 128 (which, through trial and error,
has been found to be a well-performing value).

In Figure \ref{fig:singleproc} we show the time taken for the full
Ludwig (top left) and MILC (top right) application test cases, across
the different architectures, where the testcase size (for each
application) has been chosen to fit onto a single processor. These
results include all overheads and have been run for 1000 timesteps
(Ludwig) or conjugate gradient iterations (MILC): this arbitrary
number is chosen to be large enough to be representative of production
usage (and smooth out any fluctuations), where each test is performed
on an otherwise unoccupied node.  The results in the figure are decomposed into
the main kernels, plus the remainder which includes all other sections
of the code (including PCI-express data transfer overheads in the GPU
case). 

Considering the traditional CPU results it can be seen that, overall,
the Ivy-Bridge has the performance advantage over the Haswell, where
the difference is more pronounced for Ludwig.  The AMD processor is
seen to perform significantly less well than both its Intel
counterparts. Now, when we extend our analysis to the remaining
processor types, we clearly see the advantages offered by these modern
devices.  The GPUs are between 3 and 4 times as fast as the
Ivy-Bridge, for both Ludwig and MILC. The Xeon Phi performance is seen
to be significantly better than Ivy-bridge but worse than the GPU;
better for MILC than for Ludwig.

As discussed in Section \ref{sec:targetDP}, the optimal choice of data
layout and Virtual Vector Length (VVL) varies across architectures,
and targetDP allows these options to be tuned. The above results are
for the best performing choices, details of which are included
underneath each column. The bottom graph in Figure
\ref{fig:singleproc} shows the how the performance varies with such
choices (noting the log scale).  When we set \verb+VVL+ to some value
greater than one, we are introducing explicit vectorization by
presenting grid-level parallelism at the innermost level. Conversely,
when we set \verb+VVL=1+, then we revert to the na\"{\i}ve case where
we have no explicit vectorization, and the compiler instead generates
vector instructions, where possible, from the operations conducted on
each grid point. For the CPU and Xeon Phi architectures, results are
included for \verb+VVL=N+, where, in double precision, N is naturally
4 for CPU and 8 for Xeon Phi given the instructions supported by these
architectures (256-bit AVX and 512-bit IMCI respectively). It can be
seen for the CPU results that there is no dramatic advantage from such
explicit vectorisation: the compiler can typically do a reasonable job
of finding vectorization implicitly. For the Xeon Phi, however, it is
very clear that explicit vectorization is required to get optimal
performance, otherwise the degradation is several-fold. Note that
512-bit instructions will soon propagate into future ``traditional''
CPU technologies, so these results indicate that such explicit
vectorization will soon be vital across all modern CPUs. The use of a
data layout which does not match the architecture-specific access
pattern can be very detrimental. For example, for the MILC Xeon Phi
\verb+VVL=8+ case, use of the SoA layout results in a dramatic 5-fold
slowdown overall relative to the natural AoSoA layout (and the
slowdown for AoS is more modest but still significant at 12\%). GPUs
clearly perform best with SoA data layout (which permits memory
coalescing); the wrong layout can have dramatic effects, e.g. the MILC
case is 7-fold slower with AoS. It is known that in some cases, it can
be beneficial to include explicit vectorization in GPU kernels
\citep{volkov2010better}, but we do not find any advantage in doing so
for the test cases presented in this paper. Overall, it is clear that
the specific configuration is important for performance, and the
flexibility offered by our model allows the user to experiment to find
the best options on any specific architecture.

A standard methodology for comparing observed performance to the
capability of the hardware involves the ``Roofline'' model as given in
\citep{williams2009roofline}. This uses the concept of ``Operational
Intensity'' (OI): the ratio of operations (in this case double
precision floating point operations) to bytes accessed from main
memory. The OI, in Flops/Byte, can be calculated for each
computational kernel. A similar measure (also given in Flops/Byte),
exists for each processor: the ratio of peak operations per second to
the memory bandwidth of the processor. This quantity, which gives a
measure of the balance of the processor, is known as the ``ridge
point'' in the Roofline model. Any kernel which has an OI lower than
the ridge point is limited by the memory bandwidth of the processor,
and any which has an OI higher than the ridge point is limited by the
processor's floating point capability. Table \ref{tab:procdetails}
gives peak performance values for each processor, together with memory
bandwidth measurements obtained through the standard STREAM triad
benchmark \citep{McCalpin1995}. Dividing the former by the latter, the
ridge points for the Ivy-bridge, Xeon Phi and K40 processors are
determined as 5.2, 6.4 and 7.4 Flops/Byte respectively.

The numbers given in brackets after each kernel name in Figure
\ref{fig:bandwidth} give the OI for that kernel. Since all of these
are significantly lower than the ridge point values, the Roofline
model tells us that the limiting factor is memory bandwidth (rather
than floating point capability) without exception. This means that a
reasonable method to assess performance portability is to simply to
compare the memory bandwidth obtained by each kernel to that obtained
by STREAM, on each architecture.  Note that STREAM does not report
peak bandwidth, but gives an indication of what is practically
achievable in a well-performing case, so it is possible for kernels to
slightly outperform STREAM.

Figure \ref{fig:bandwidth} shows the memory bandwidth of each kernel
expressed as a percentage of STREAM. For real, complex applications,
it is over-ambitious to expect 100\% across all kernels. The
performance will depend on the specific complexities of each kernel,
but it is clear that, for a well-performing code, most kernels should
be achieving a reasonable percentage of STREAM bandwidth. First
considering the K40 GPU results in the Figure, it can be seen that all
kernels are near 100\%, except for ``Chemical Stress''. Further
investigation reveals that this kernel achieves relatively low
occupancy on the GPU because of the register usage required by each
thread in efficiently storing temporary structures. Future work will
attempt to redesign the kernel at the algorithmic level with the aim
of improving performance across all architectures.

We derive estimates of corresponding bandwidths on the Ivy-bridge and
Xeon Phi using the measured K40 bandwidth results scaled by ratios of
timing results (due to the lack of appropriate tools, on the platforms
in use, that can provide the necessary kernel-level
information). These are only estimates because they assume that the
same amount of data is being loaded from main memory across the three
architectures, which may be an oversimplification due to differing
caching architectures and policies, but nevertheless give us useful
indications. The CPU and Xeon Phi results are seen to lack the
consistency of the GPU results, more so for Ludwig (for which the
kernels are typically more complex) than MILC. Our estimation
technique may play a role in this variability, but is is also possible
that the GPU architecture, designed for data throughput, is more
effective in achieving good bandwidth for such real
applications. Nevertheless, for the CPU, we see 60\% or higher of STREAM
bandwidth for all but two kernels. The variability for the Xeon Phi is
higher, but this is not surprising since the current product is known
to have architectural limitations which are being addressed by a
complete re-design for the next-generation ``Knights Landing''
product. Since we are already carefully tuning the data layout and
corresponding access patterns, it is not clear that there are any
major improvements which could be made to the application kernels, but
it is possible that tuning of more subtle system settings may be
advantageous and we will explore this in future work. Intel quote
80GB/s (around 50\% of STREAM) as a threshold on the current Xeon Phi,
above which memory bandwidth can be considered as being good
\citep{XeonPhiTuning}, and we are achieving this for 9 of our 13
kernels.

\section{Scaling}\label{sec:scaling}

\begin{table}
\footnotesize
\centering
\begin{tabular}{|c|c|c|} \hline
    & Titan& ARCHER \\ \hline  \hline
Location&Oak Ridge&University of \\ 
        &National Laboratory&Edinburgh \\ \hline 
Product&Cray XK7&Cray XC30 \\ \hline
Per Node&One Interlagos \&  & Two Ivy-Bridge  \\ 
& one K20X&\\ \hline
Nodes&18,688&4,920\\ \hline
Interconnect&Cray Gemini&Cray Aries\\ 
\hline\end{tabular}
\normalsize
\caption{Technical details of the supercomputers used in the scaling analysis. See Table \ref{tab:procdetails} for details of the processors.}\label{tab:scdetails}
\end{table}

In this paper we have presented our targetDP model, and demonstrated
its effectiveness in allowing the same application source code to
perform well across modern architectures. The model is relevant for
intra-node parallelism, so for use on multi-node systems it must be
combined with a higher-level paradigm. Both the applications we have
presented use MPI for inter-node communications. In this section, we
analyse the performance of our MPI+targetDP applications at scale.

\begin{figure*}[t]
\vspace{0cm}\centerline{\includegraphics[width=6cm]{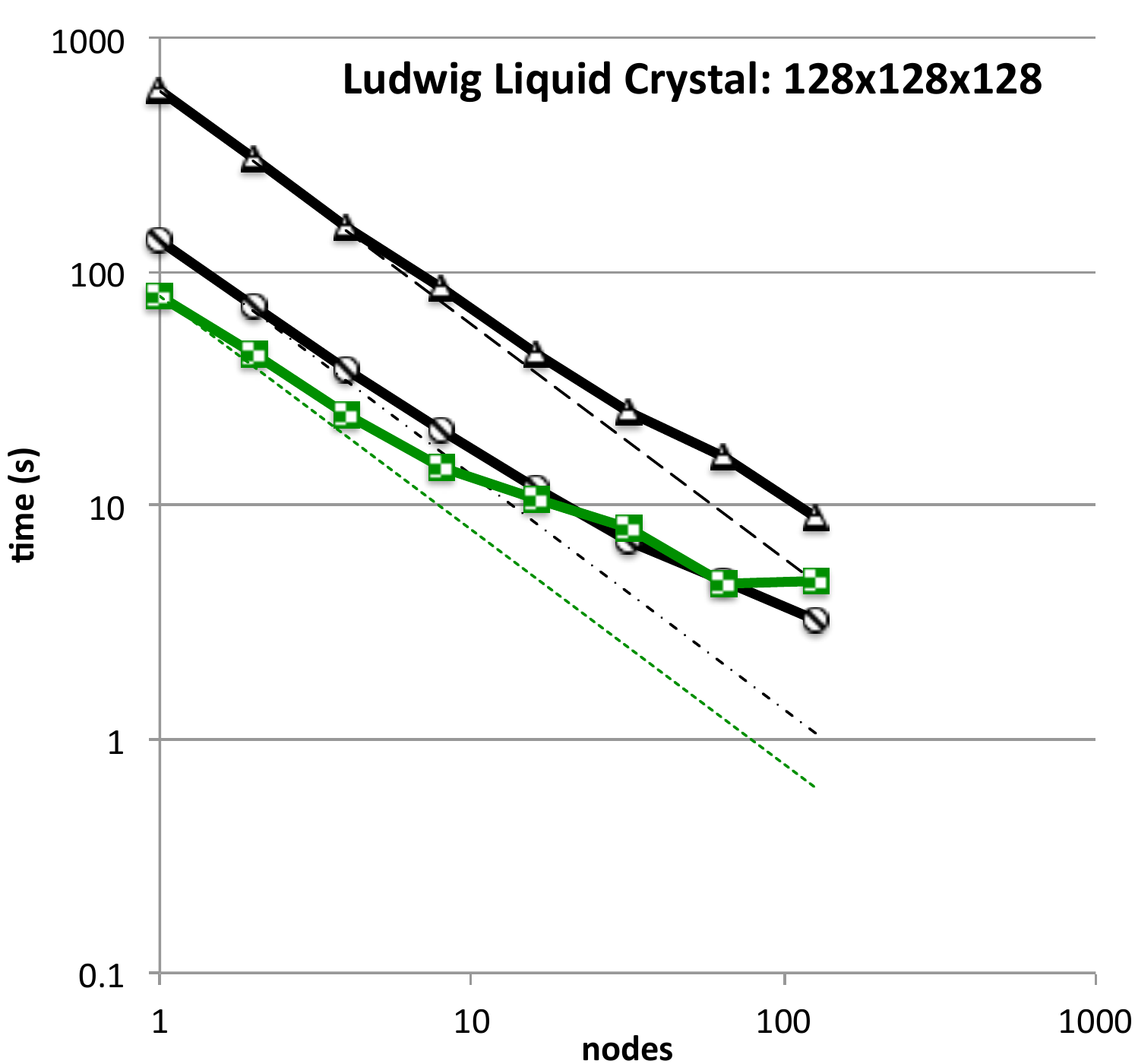}
\hspace{2.8cm}
\includegraphics[width=6cm]{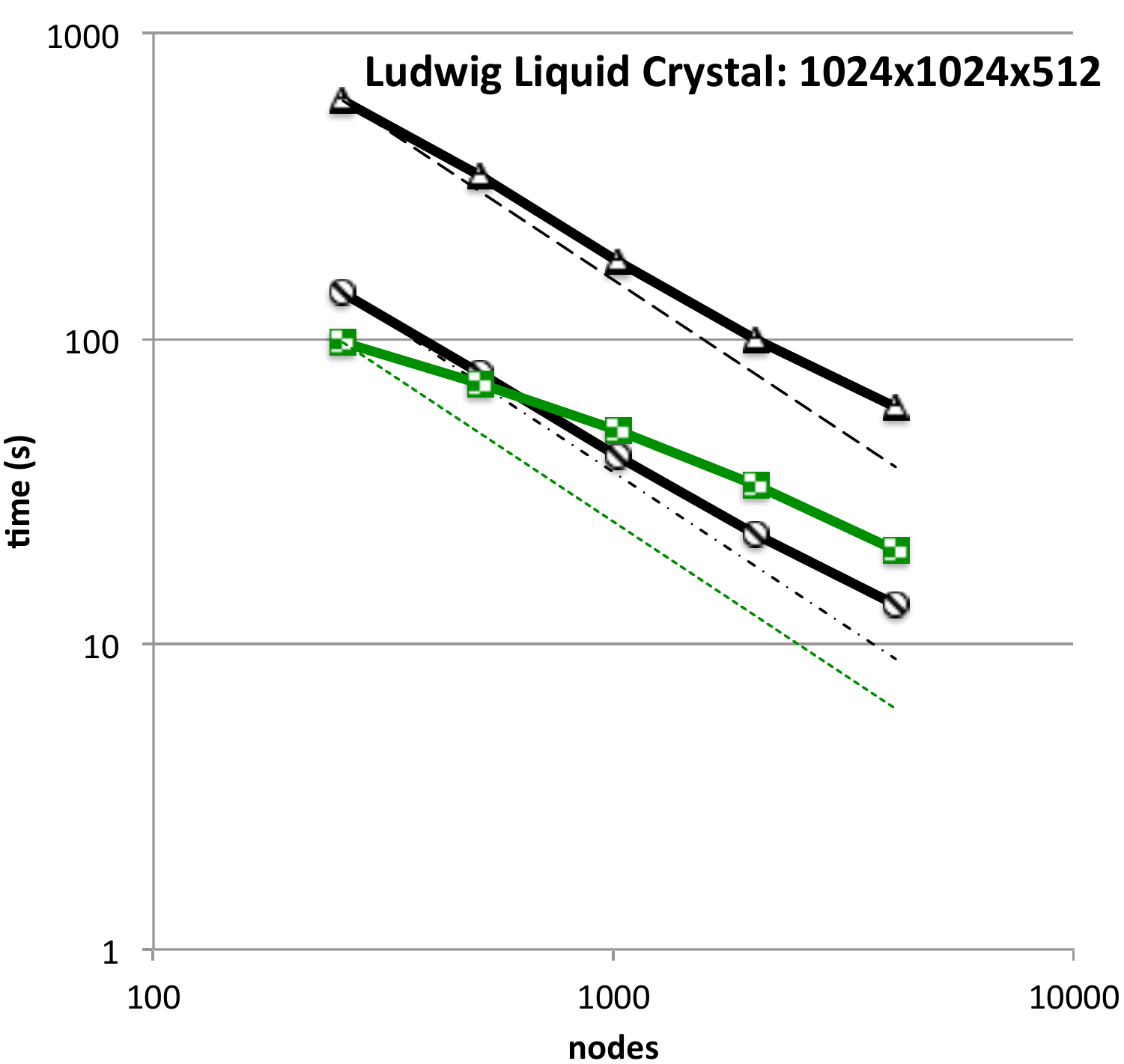}}
\vspace{0cm}\centerline{\includegraphics[width=6cm]{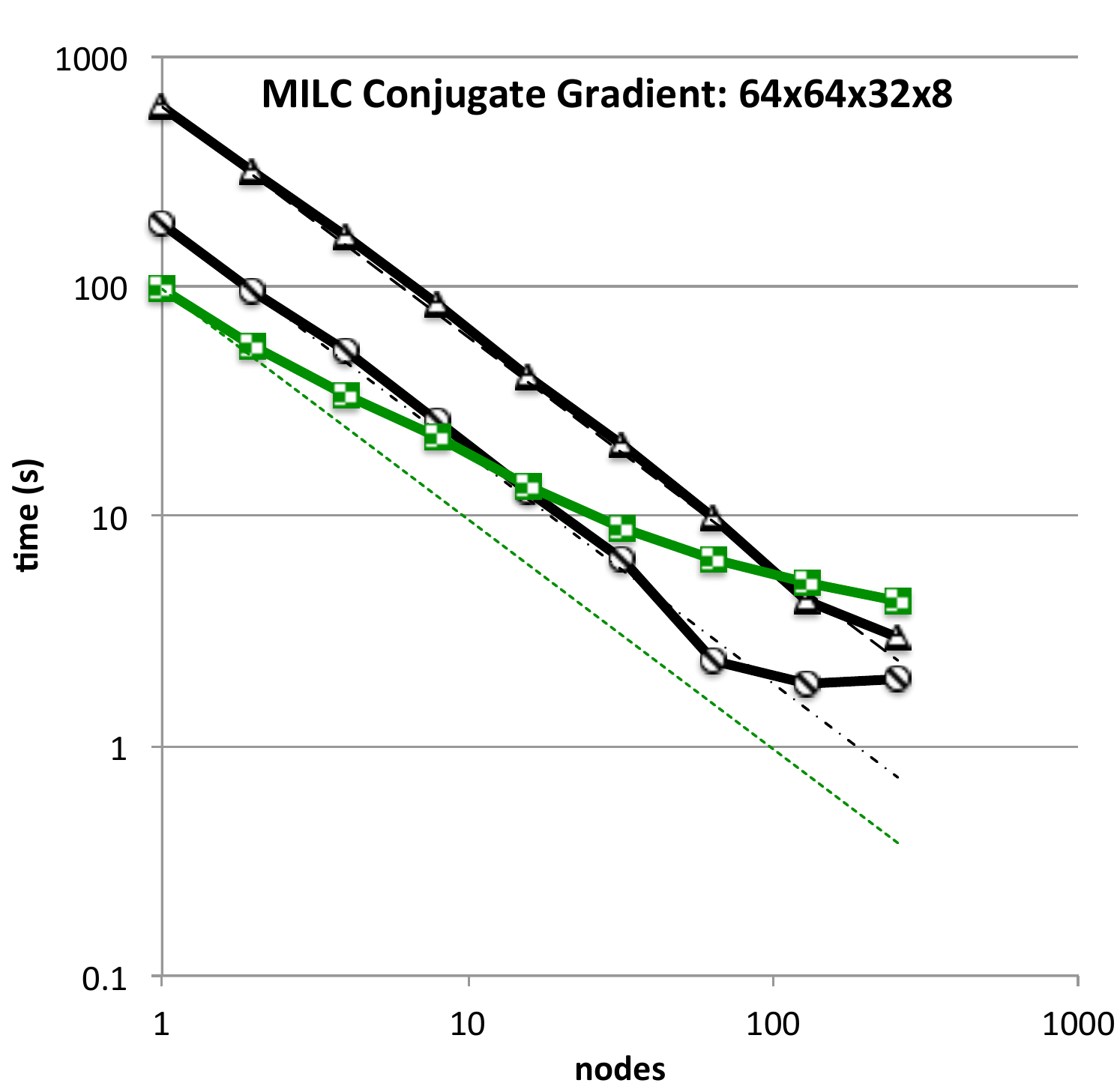}
\includegraphics[width=3cm]{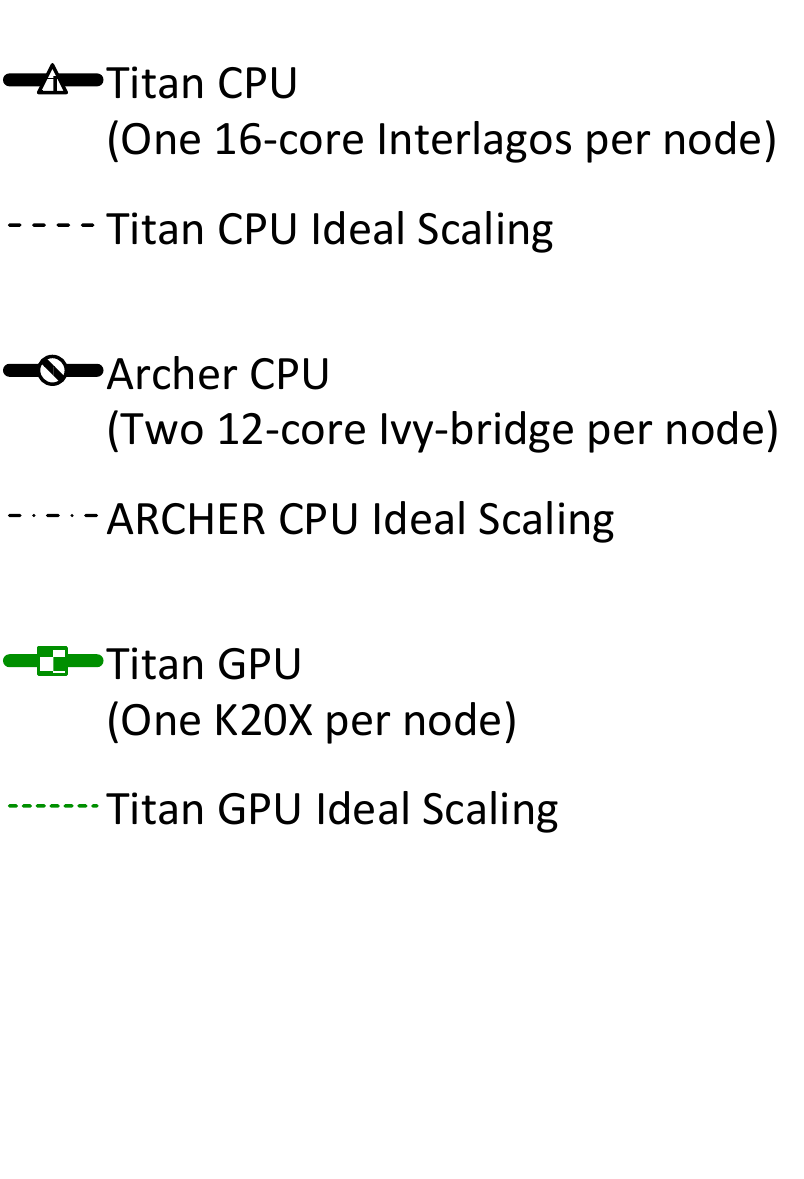}
\includegraphics[width=6cm]{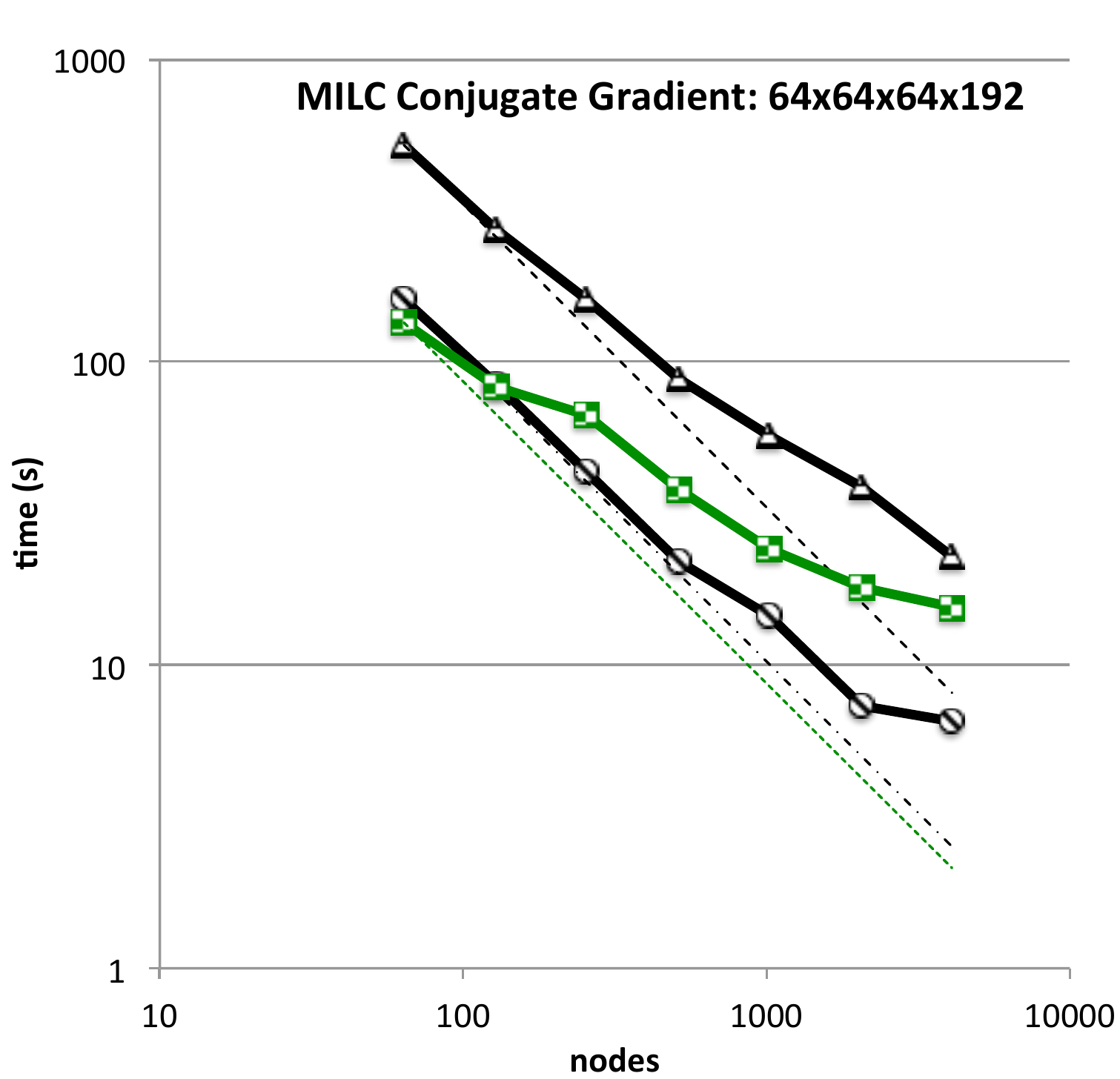}}
\vspace{-0.3cm}\caption{The strong scaling, on Titan and ARCHER, of Ludwig (top) and MILC (bottom) for small (left) and large (right) problem sizes. For
  ARCHER, both CPUs are used per node. For Titan, we include results
  with and without GPU utilization. }\vspace{-0.2cm}\label{fig:scaling}
\end{figure*}

For this analysis we use the Titan and ARCHER supercomputers, for
which details are given in Table \ref{tab:scdetails}. On each node,
Titan has one 16-core Interlagos CPU and one K20X GPU, whereas ARCHER
has two 12-core Ivy-bridge CPUs. In this section, we evaluate on a
node-by-node basis. For Titan, a single MPI task per node, operating
on the CPU, is used to drive the GPU on that node. We also include,
for Titan, results just using the CPU on each node without any
involvement from the GPU, for comparison. This means that, on a single
node, our Titan results will be the same as those K20X and Interlagos
results presented in the previous section (for the same test case). On
ARCHER, however, we fully utilize both the processors per node: to do
this we use two MPI tasks per node, each with 12 OpenMP threads (via
targetDP). So the single node results for ARCHER are twice as fast as
those Ivy-bridge single-processor results presented in the previous
section.  Note that targetDP has no capability to decompose data
parallel operations across both CPUs and GPUs within a single run. The
advantage of this would be minimal, compared to sole use of the GPU,
even if good load balancing was achieved, since the memory bandwidth
on the GPU is so much higher than the CPU, so we do not believe it is
worth the substantial complexity that would be required. Similar to
the previous section, we run for 1000 timesteps or iterations and,
since these runs are across multiple nodes on a shared system, we
reduce the effects of any interference from other jobs by performing
each run several times and select the minimum timing result.

In Figure \ref{fig:scaling}, we show strong scaling, for two different
problem sizes, for both Ludwig and MILC. The small problem sizes on
the left are those which were used in our single-processor analysis in
the previous section. The problem sizes on the right are chosen to be
representative of current problems requiring HPC.  For each, we see
how the time to solution deceases as we increase the number of
nodes. 

For the small problem cases, on the CPUs, the scaling is seen to be
excellent up to around 32 nodes. After that point, the local
problem size becomes too small and communication dominates. For the
GPU case, strong scaling is still observed up to this point (i.e. the
time to solution continues to decrease), but the scaling is seen to
deviate from the ideal case. At low node counts the single GPU per
node on Titan outperforms the two CPUs per node on ARCHER, but there
is a crossover point (around 32 nodes) after which the ARCHER
performance is better.

This behaviour is expected to be attributable to the extra data
transfers across the PCI-express bus necessary when exchanging halo
data between GPUs (via the host CPUs). We expect that hardware
improvements in future systems will reduce these overheads: most
notably the introduction of the higher bandwidth and lower latency
NVLINK as a high performing replacement for PCI-express in future
NVIDIA models \citep{NVLINK}. There are also potential improvements in
software through use of CUDA-aware MPI, which should reduce overheads
through removal of unnecessary data buffering and CPU
involvement. \revised{This can be used in a straightforward manner by passing pointers to the target rather than host copies of the data structures directly into MPI calls.} There already exists such a library on Titan, but it
still lacks maturity and is not used here because of incompatibilities
with the specific communication patterns used in these applications. 

For the larger problem sizes, there exists a minimum number of nodes
for which the problem can fit into total memory: 256 nodes for Ludwig
and 64 nodes for MILC (i.e. the Ludwig problem size is larger).  It
can be seen that the behaviour is similar, with the crossover point
dependent on the problem size, occurring at around 512 nodes for Ludwig
and 256 nodes for MILC.

Overall, these results confirm that targetDP can be combined
effectively with MPI for use on large-scale supercomputers for these
types of problem.

\section{Conclusions}
Large-scale HPC systems are increasingly reliant on highly parallel
processors such as NVIDIA GPUs and Intel Xeon Phi many-core CPUs to
deliver performance. It is unsustainable for the programmer to write
separate code for each of these different architectures (and also
traditional CPUs). In this paper we presented a pragmatic solution to
this problem, for grid-based applications. The targetDP model consists
of a very lightweight framework, which abstracts either OpenMP (for
multi-core or many-core CPUs) or CUDA (for NVIDIA GPUs), whilst
allowing good vectorization on the former. We showed that use of this
approach can achieve real performance portability across the
architectures for complex applications, and that it can effectively be
combined with MPI for use on large supercomputers. An important
finding is that the explicit vectorization capabilities of the model
are vital in order to make efficient use of the current Xeon Phi
processors which feature 512-bit SIMD units, and are expected also be
vital on imminent mainstream CPU products, for which the width of the
vector instruction will increase to this size.

Our work, for which the software is open source and freely available
\citep{targetDPweb}, has so far been limited to structured grid
applications, but the model or ideas may well be of interest more
widely. The main architecture that we have not yet addressed is the
AMD GPU, which we should be able to target in a straightforward manner
through development of an OpenCL implementation of targetDP (that
largely mirrors the CUDA implementation). It should be similarly
straightforward to create Fortran implementations (for the key
architectures discussed in this paper) to support the many HPC
grid-based applications written in that language. We also look forward
to imminent hardware advances from Intel and NVIDIA: the next
generation Xeon Phi and GPU products will both exploit high-bandwidth
stacked memory, which should significantly boost performance of our
memory-bandwidth bound applications. Furthermore, the new low latency
and high bandwidth NVLINK CPU to GPU interconnect promises to reduce
the communication overheads in strong scaling on GPU systems.


\section*{Acknowledgments}
\revised{ AG acknowledges support from the embedded CSE programme of the ARCHER UK National Supercomputing Service (http://www.archer.ac.uk)} and the
European Union funded PRACE programme. KS acknowledges support from
United Kingdom EPSRC grant EP/J007404.  This research used resources
of the Oak Ridge Leadership Computing Facility at the Oak Ridge
National Laboratory, which is supported by the Office of Science of
the U.S. Department of Energy under Contract No. DE-AC05-00OR22725,
and the United Kingdom EPSRC funded ARCHER service. We thank Daniel
Holmes and Arno Proeme for providing constructive feedback on this
paper.



%
\bibliographystyle{apa}
\bibliography{paper}

\end{document}